\documentclass[aps,preprint,amsmath,amssymb,t1,11pt]{revtex4}
\usepackage{graphicx}
\usepackage{epstopdf}
\usepackage[T1]{fontenc}
\usepackage{xcolor}
\usepackage{slashed}
\usepackage{multirow}
\usepackage{slashed}

\begin{document}
\draft

\title{Search for new physics effects in $\nu\bar{\nu}\gamma$ production at a Tera-Z factory }

\author{H. Denizli}
\email[]{denizli_h@ibu.edu.tr}
\affiliation{Department of Physics, Bolu Abant Izzet Baysal University, 14280, Bolu, T\"{u}rkiye.}

\author{A. Senol}
\email[]{senol_a@ibu.edu.tr}
\affiliation{Department of Physics, Bolu Abant Izzet Baysal University, 14280, Bolu, T\"{u}rkiye.}

\author{M. K{\"o}ksal}
\email[]{mkoksal@cumhuriyet.edu.tr}
\affiliation{Department of Physics, Sivas Cumhuriyet University, 58140, Sivas, T\"{u}rkiye.}
\affiliation{Sivas Cumhuriyet University Nanophotonics Application and Research Center (CUNAM), Sivas, 58140, T\"{u}rkiye.}

\date{\today}

\begin{abstract}
Rare decays of the Z boson provide a sensitive probe for physics beyond the standard model. This study investigates the $e^{+}e^{-} \to Z \to \nu\bar{\nu}\gamma$ process within the context of the Tera-Z programmes at future colliders such as the FCC-ee and CEPC. The SM predicts a one-loop branching ratio of $7.16 \times 10^{-10}$ for $Z \to \nu\bar{\nu}\gamma$, a value four times smaller than the current experimental limit from the LEP. To explore this window for new physics, we parameterize anomalous $Z\nu\bar{\nu}\gamma$ interactions using an Effective Field Theory framework, considering both dimension-6 and dimension-8 operators.
A detailed simulation is performed by generating signal and background events with MadGraph, modeling particle showers with Pythia, and simulating detector effects with Delphes. The analysis employs key kinematic variables—including the photon energy ($E_\gamma$), missing transverse energy ($\slashed{E}_T$), and the missing transverse energy significance ($S_{\slashed{E}_T}$) to isolate the signal. The results yield upper limits on the anomalous couplings, from which we infer branching ratios for $Z \to \nu\bar{\nu}\gamma$ on the order of $10^{-9}$. This represents a significant improvement of several orders of magnitude over the LEP sensitivity. Consequently, this study demonstrates the unique potential of the Tera-Z runs not only to test the SM loop-level predictions with unprecedented precision but also to tightly constrain or reveal new anomalous interactions.

\end{abstract}

\maketitle

\section{Introduction}

The search for beyond the Standard Model (BSM) requires a comprehensive and precise approach, as the properties and mass ranges of potential new particles remain largely unknown. This challenge demands an exploration strategy that combines enhanced measurement precision, sensitivity to rare processes, and, ultimately, higher energy capabilities. Future electron-positron colliders with high luminosity are uniquely positioned to address these demands effectively. In the pristine environment of an electron-positron collider, a multifaceted program of precision measurements, searches for new particles, and investigations into rare or forbidden processes will provide access to high-energy scales through loop diagrams and probe tiny couplings via mixing or forbidden interactions. This program also opens avenues for the direct discovery of dark matter, heavy neutrinos, and other weakly interacting particles. The remarkable agreement between the observed masses of the top quark and Higgs boson with the predictions of the minimal SM is highly non-trivial and imposes stringent constraints on the nature of potential new physics. Any quantum effects from new physics must be smaller than the current experimental uncertainties, making it essential to reduce these uncertainties to levels that can reveal such subtle contributions. Furthermore, the reduction in experimental uncertainties will significantly enhance our understanding of the SM's fundamental parameters, including the strong, weak, and electromagnetic coupling constants at the weak scale, with precision improvements of up to two orders of magnitude.

Rare Z decays offer a distinctive perspective on the potential for BSM. These decays, which are ordinarily suppressed within the SM, can be substantially amplified by new physics effects, rendering them sensitive probes for phenomena not accounted for by the SM. For example, the anticipated sensitivity of future colliders such as the giga-Z linear collider is approximately ($10^{-8}$), underscoring the necessity of detecting any new physics effects that surpass this threshold \cite{Perez:2003ad}. The presence of non-universal Z bosons has been demonstrated to enhance the branching ratios of flavor-changing neutral current (FCNC) mediated decays, such as $Z \to b\bar{s}$, by a factor of at least one order of magnitude \cite{Mohanta:2005gm}. This observation underscores the utility of these decays as indicators of new physics. Furthermore, rare decays involving double heavy quarkonia, such as $Z \to J/\Psi J/\Psi$, can be significantly enhanced due to electromagnetic transitions, suggesting that these processes could serve as effective probes for new physics \cite{Gao:2022mwa}. Moreover, the LEP bounds on rare Z decays, such as $Z \to \gamma\gamma\gamma$, $Z \to \nu\bar{\nu}\gamma \gamma$ and $Z \to \nu\bar{\nu}\gamma$ have been crucial in constraining new physics scenarios, including interactions involving neutrinos and photons, as well as potential light pseudoscalars \cite{Perez:2003ad}. Consequently, the ongoing investigation of rare Z decays remains a promising avenue for the unveiling of new physics. Future advances in experimental techniques are expected to provide critical tests of the SM at the loop level.

The Tera-Z phase of future circular electron-positron colliders will play a unique and pivotal role in the advance of electroweak precision physics \cite{Blondel:2018mad}. By producing an enormous number of on-shell Z bosons, the Tera-Z phase will achieve unparalleled precision in measuring Z boson properties and electroweak precision observables \cite{Maura:2024zxz}. The data collected during this phase will be instrumental in rigorously testing the internal consistency of the SM and could potentially reveal deviations indicative of new physics. This exceptional sensitivity makes the Tera-Z phase a cornerstone for probing the SM with unprecedented accuracy and exploring the potential existence of physics BSM \cite{Allwicher:2024sso,Bao:2025tqs}.

Two leading candidates for achieving the Tera-Z phase are the Circular Electron Positron Collider (CEPC) \cite{CEPCStudyGroup:2025kmw} and the electron-positron option of the Future Circular Collider (FCC-ee) \cite{FCC:2018evy,FCC:2025lpp,Altmann:2025feg}. 
While both projects are primarily envisioned as future Higgs factories, they also offer exceptional opportunities for flavor physics, extending their capabilities beyond the standard Tera-Z phase. The CEPC, with a center-of-mass energy of 91.2 GeV, is initially designed to operate at the Z-pole with high luminosity ($\mathcal{L}_{int}$=100 ab$^{-1}$), with the objective of collecting $0.7 \times 10^{12}$ Z bosons \cite{CEPCPhysicsStudyGroup:2022uwl}. Recent design updates \cite{CEPCStudyGroup:2023quu} indicate that the CEPC could now deliver over $3 \times 10^{12}$ Z bosons within a two-year period. 

The proposed Z-pole operation plan of FCC-ee comprises data-taking at three center-of-mass energies: 87.9 GeV, 91.2 GeV, and 94.3 GeV. These energies are chosen to correspond to half-integer spin tunes, thereby enabling precise energy calibration through resonant depolarization. Approximately half of the total dataset will be collected at 91.2 GeV. The integrated luminosity is expected to reach $2 \times 10^{5}$ times that achieved by the LEP at the Z pole, providing a sample of about $5 \times 10^{12}$ Z bosons for detailed studies \cite{FCC:2025lpp}. Based on the current operational scenarios, the FCC-ee is anticipated to deliver integrated luminosities of approximately $40~\text{ab}^{-1}$ at $\sqrt{s}=87.9~\text{GeV}$, $125~\text{ab}^{-1}$ at the $Z$-pole energy $\sqrt{s}=91.2~\text{GeV}$, and $40~\text{ab}^{-1}$ at $\sqrt{s}=94.3~\text{GeV}$.

The search for BSM has often focused on rare or suppressed processes, such as the single-photon decay $Z\to X+\gamma$, where $X$ shows any neutral, invisible state. This decay has played a significant role in probing potential new physics, particularly through studies conducted at the LEP. These experiments near the Z pole set an upper limit on the $BR(Z \to\nu\bar{\nu}\gamma)<10^{-6}$ \cite{L3:1997exg,DELPHI:1996drf}, which is significantly higher than the SM prediction of $BR(Z \to\nu\bar{\nu}\gamma)=7.16 \times 10^{-10}$ \cite{Hernandez:1999xn}. The SM contribution arises primarily from one-loop diagrams, including neutrino magnetic dipole transitions and box diagrams. For this reason, the large discrepancy between the experimental limit and the SM prediction leaves a promising window for exploring new physics effects in single-photon decays of the Z boson. Consequently, the electromagnetic interaction between the Z boson and neutrinos can be explored for the new physics research \cite{Hernandez:1999xn, Perez:2003ad}.

\begin{figure}[!ht]
\includegraphics[scale=1.0]{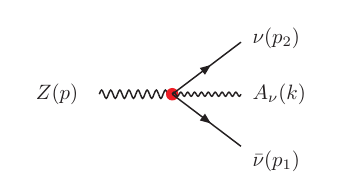}
\caption{Representation of $Z\nu \bar{\nu}\gamma$ vertex (red dot) in the Effective
Lagrangian approach. \label{vertex}}
\end{figure}

A well-established framework to explore possible deviations from the SM is the Effective Field Theory (EFT) approach. Within this framework, non-standard neutrino–photon interactions can be systematically studied through higher-dimensional operators that parameterize new physics in a model-independent way. In particular, the non-standard $Z\nu\bar{\nu}\gamma$ coupling can be described by a dimension-6 operator, which represents the lowest order at which such an interaction can appear. This operator is expressed as follows~\cite{Dusedau:1986xq}:
\begin{eqnarray}
\label{eq.1}
\mathcal{L}=\frac{e}{\Lambda^{2}}(\kappa_{1}{F}_{\mu\nu}+\kappa_{2}\tilde{{F}}_{\mu\nu}) \bar{\Psi}
\gamma^{\mu}(a+b\gamma_5)\Psi
Z^{\nu}
\end{eqnarray}
where $\Lambda$ is the new physics scale, $\tilde{F}_{\mu\nu} = \frac{1}{2}\epsilon_{\mu\nu\rho\sigma}F^{\rho\sigma}$ and $\kappa_1 (\kappa_2)$ is CP-odd (even) dimension-6 couplings. We set $a^2+b^2=1/2$.
In addition to dimension-6 operator, $Z\nu \bar{\nu}\gamma$ vertex arises from the dimension-8 operators without obscuration by dimension-4 SM contributions and dimension-6 SMEFT effects given as follows \cite{Maya:1998ee}
\begin{equation} 
O^{8}_{1}={\mathrm{i}}
(\phi^{\dagger} \phi)\bar{\ell}^{a}_{L}
\tau^{i} \gamma^{\mu}
D^{\nu}\ell^{a}_{L}W^{i}_{ \mu \nu},
\label{eiop1} 
\end{equation}
\begin{equation} 
O^{8}_{2}= {\mathrm{i}}
(\phi^{\dagger} \phi)\bar{\ell}^{a}_{L}
\gamma^{\mu} D^{\nu} \ell^{a}_{L} B_{\mu \nu},
\label{eiop2} 
\end{equation}
\begin{equation}
O^{8}_{3}= {\mathrm{i}} (\phi^{\dagger}
D^{\mu} \phi)\bar{\ell}^{a}_{L} \gamma^{\nu}
\tau^{i} \ell^{a}_{L}W^{i}_{\mu \nu},
\label{eiop3}
\end{equation}
\begin{equation}
O^{8}_{4}= {\mathrm{i}} (\phi^{\dagger} D ^{\mu}
\phi)\bar{\ell}^{a}_{L}
\gamma^{\nu}
\ell^{a}_{L} B_{\mu \nu}. \label{eiop4}
\end{equation}
Here, $W^{i}_{\mu \nu}$ and $B_{\mu \nu}$ show $\mathrm{SU(2)_{L}}$ and $ \mathrm{U(1)_{Y}}$
tensor field strength tensors respectively, as well as the  $\ell^{a}_{L}$ represents $\mathrm{SU(2)_{L}}$ left-handed lepton doublet, $\tilde{\phi}=\mathrm{i} \tau^{2} \phi^{\ast}$ is the Higgs field, $\tau_{i}$ are the Pauli matrices and $D_{\mu}$ is covariant derivative. Using the kinematic variables shown in Fig.\ref{vertex}, following spontaneous symmetry breaking, the effective operators Eqs.(\ref{eiop1})-(\ref{eiop4}) give rise to the following parametrization of the $Z\nu \bar{\nu}\gamma$ effective couplings;
\begin{eqnarray}
\label{eq.2}
\mathcal{L}_{\mu\nu}=\frac{\epsilon_8}{v^{2}}\bar u(p_2)(1-\gamma_5)(k_{\mu}\gamma_{\nu}-\not kg_{\mu\nu})v(p_1)
\end{eqnarray}
where $v$ is the vacuum expectation value. The coefficients $\epsilon_8$ summarize all the information that can be gathered from the heavy degrees of freedom associated with new physics effects and are expressed in terms of dimensionless coupling constants $\alpha_i$ and the scale $\Lambda$: $\epsilon_8 =\epsilon_8^2+\epsilon_8^3+\epsilon_8^4$ with $\epsilon_8^i=\alpha_8^i(v/\Lambda)^4$ (where $i=2,3,4$). Here, the coefficient $\epsilon^{1}_{8}$ is omitted from the expression because the corresponding operator in Eqs. (\ref{eiop1})-(\ref{eiop4}) does not generate a $Z\nu\nu\gamma$ vertex after electroweak symmetry breaking.

Within this framework, we first illustrate the parametric dependence of the anomalous $Z\nu\bar{\nu}\gamma$ vertex on the effective operator coefficients by examining the partial decay width $\Gamma(Z \to \nu\bar{\nu}\gamma)$ under controlled kinematic conditions.  The differential decay width $Z\to\nu \bar{\nu}\gamma$ can be obtained as a function of both $\kappa_1/\Lambda^2$, $\kappa_2/\Lambda^2$ and $\alpha_8/\Lambda^4$ coupling using the effective vertex defined in Eq.(\ref{eq.1}) and Eq.(\ref{eq.2}) as follows:
\begin{equation} 
\frac{d\Gamma}{dx}=\frac{\alpha}{192\pi^2}(\frac{\kappa_1^2+\kappa_2^2}{\Lambda^4})M_Z^5(\frac{2x^3}{3}-\frac{x^4}{3})
\label{dgdx1} 
\end{equation}
and 
\begin{equation} 
\frac{d\Gamma}{dx}=\frac{v^4}{18\pi^2}(\frac{\alpha_8^2}{\Lambda^8})M_Z^5(x^3-x^4)
\label{dgdx2} 
\end{equation}
where $x$ is proportional to the photon energy and is defined as $x=2E_{\gamma}/M_Z$. To obtain the decay width of the $Z\to\nu \bar{\nu}\gamma$ decay, we integrate the differential decay expressions given in Eqs. (\ref{dgdx1}) and (\ref{dgdx2}) as follows;
\begin{equation}
    \Gamma(Z\to\nu \bar{\nu}\gamma)=\int^1_{2E_{\gamma}^{min}/M_Z}\frac{d\Gamma}{dx}dx
\end{equation}

\begin{figure}[!htbp]
\includegraphics[scale=0.50]{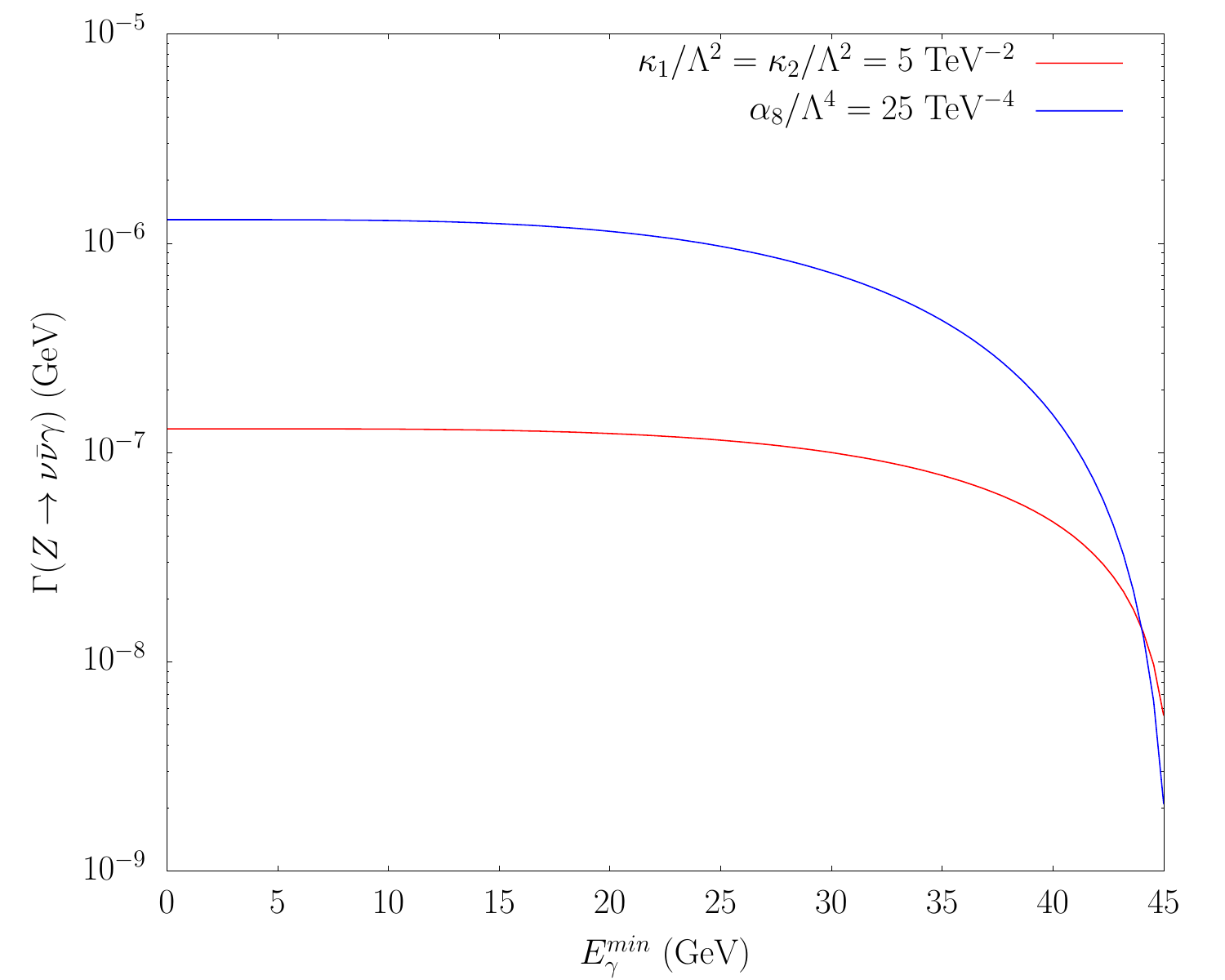}
\caption{Partial decay width $\Gamma(Z \to \nu\bar{\nu}\gamma)$ as a function of the minimum photon energy $E_{\gamma}^{\min}$ for representative benchmark values of the effective operator coefficients. The solid blue curve corresponds to the dimension-6 contribution with $\kappa_{1}/\Lambda^{2} = \kappa_{2}/\Lambda^{2} = 5~\mathrm{TeV}^{-2}$, while the solid orange curve shows the dimension-8 contribution with 
$\alpha_{8}/\Lambda^{4} = 25~\mathrm{TeV}^{-4}$. \label{Fig.2}}.
\end{figure}

Fig.~\ref{Fig.2} shows the predicted behavior of the obtained partial decay width $\Gamma(Z \to \nu \bar{\nu} \gamma)$ as a function of the minimum photon energy $E_{\gamma}^{\min}$ for two different sets of effective operator coefficients. 
The two curves correspond to the contributions of dimension-6 operators (with $\kappa_{1}/\Lambda^{2} = \kappa_{2}/\Lambda^{2} = 5~\text{TeV}^{-2}$) and a dimension-8 operator (with $\alpha_{8}/\Lambda^{4} = 25~\text{TeV}^{-4}$). 
As shown in the figure, the decay width decreases monotonically as $E_{\gamma}^{\min}$ increases. This trend is a natural consequence of the phase-space suppression encoded in the integrands of Eqs.~(\ref{dgdx1}) and (\ref{dgdx2}), which contain factors such as 
$2x^{3}/3 - x^{4}/3$ (dimension-6) and $x^{3} - x^{4}$ (dimension-8). Imposing a larger value of $E_{\gamma}^{\min}$ restricts the integration range in $x$, reducing the accessible phase space and thus lowering the resulting partial width. This simplified setup allows for a transparent interpretation of the photon energy dependence before turning to the detailed collider-level analysis.

In this work, we study the impact of the dimension-6 and dimension-8 operators on the $Z\nu\bar{\nu}\gamma$ vertex by first analyzing the partial decay width $\Gamma(Z\to\nu\bar{\nu}\gamma)$ as an isolated decay process, and then relating these results to the full production process $e^{+}e^{-}\to\nu\bar{\nu}\gamma$, which is the 
relevant channel at a Tera-$Z$ collider and is dominated by on-shell $Z$-boson production. We emphasize that the decay study is used only to illustrate the dependence of the vertex on the EFT couplings, while the phenomenological predictions 
for collider rates are based on the full $e^{+}e^{}\to\nu\bar{\nu}\gamma$ process.

For this purpose, the operators defined in Eqs.~(\ref{eq.1})-(\ref{eq.2}) are implemented into a Universal FeynRules Output (UFO)~\cite{Degrande:2011ua} model, generated with the \texttt{FeynRules} package \cite{Alloul:2013bka}, and subsequently employed in \texttt{MadGraph5\_aMC@NLO v3.5.6}~\cite{Alwall:2014hca}. Within this framework, we simulate both the signal and background events for the $e^+e^- \to \nu \bar{\nu}\gamma$ process. A detailed phenomenological analysis is then carried out by scanning the relevant parameter space of the anomalous couplings. This approach allows us to quantify the sensitivity of the BR($Z \to \nu \bar{\nu}\gamma$) to the effective operators and to identify the regions where deviations from the SM predictions become most significant.
\begin{figure}[htbp]
\includegraphics[scale=0.5]{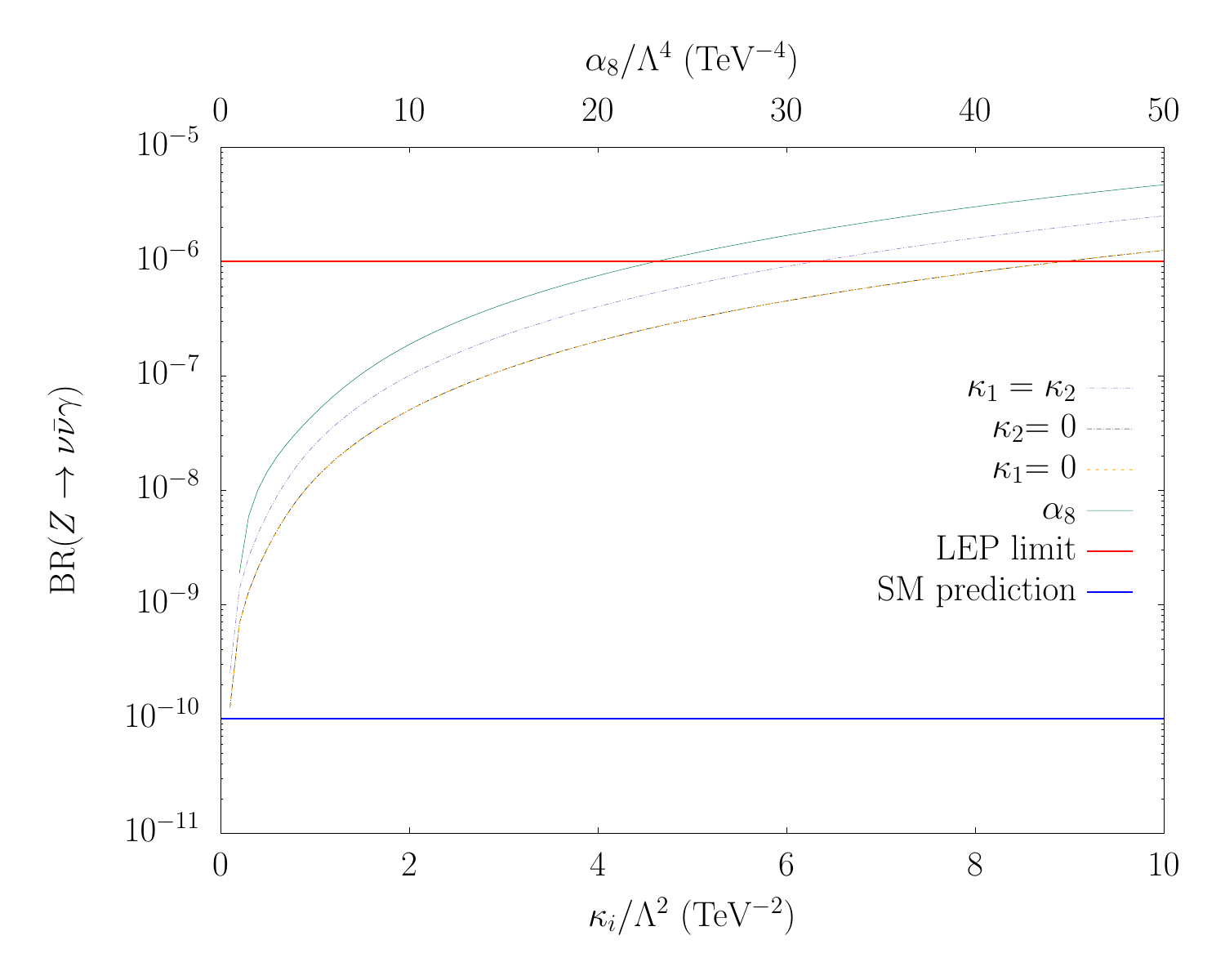}
\caption{BR of the $Z \to \nu \bar{\nu}\gamma$ decay as a function of dimension-6 ($\kappa_1/\Lambda^2$ and $\kappa_2/\Lambda^2$) and dimension-8 ($\alpha_8/\Lambda^4$) couplings. \label{Fig.3}}.
\end{figure}
Fig. \ref{Fig.3} illustrates BR($Z \to \nu \bar{\nu}\gamma$) decay as a function of both dimension-6 ($\kappa_1/\Lambda^2$ and $\kappa_2/\Lambda^2$) and dimension-8 ($\alpha_8/\Lambda^4$) anomalous couplings. The figure also displays the SM prediction and the experimental limit set by the LEP experiments \cite{L3:1997exg,DELPHI:1996drf}.
Within the SM, the decay probability is highly suppressed. However, the introduction of dimension-6 and dimension-8 operators results in a significant enhancement of the BR.
The analysis involves the comparison of different coupling scenarios, including $\kappa_1/\Lambda^2= \kappa_2/\Lambda^2$ (two dot dashed purple line), $\kappa_1/\Lambda^2$ (dashed orange line), and $\kappa_2/\Lambda^2$ (dot-dashed black line), with experimental constraints. As shown from figure, if $\kappa_1/\Lambda^2$ and $\kappa_2/\Lambda^2$ couplings are both non-zero, the contributions add up and the decay rate shows the fastest increase. However, when one coupling is zero while the other is varying, the increase is slower than in isolation. On the other hand, BR for $\alpha_8/\Lambda^4$ is at a lower level compared to $\kappa_1/\Lambda^2= \kappa_2/\Lambda^2$ interactions of the same strength. The results of this analysis indicate that the FCC-ee and the CEPC could test such anomalous interactions, thereby providing potential insights into BSM.
\section{Analysis procedure of $\nu\bar{\nu}\gamma$ production at THE FCC-ee/CEPC }
\begin{figure}[htbp]
    \centering
    \includegraphics[width=0.85\textwidth]{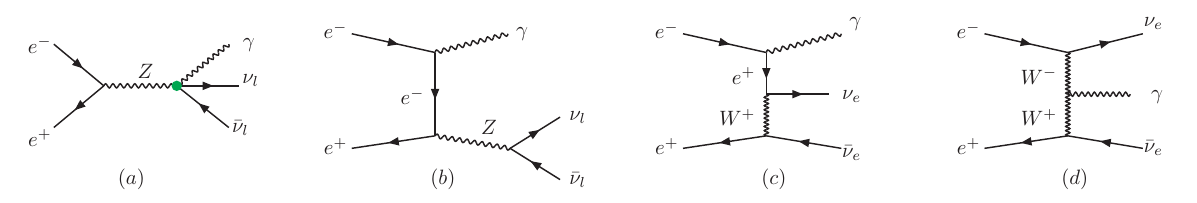}
    \caption{Representative Feynman diagrams for the $e^+e^- \to \nu \bar{\nu}\gamma$ process: (a) represents the new physics signal contribution, while diagrams~(b)–(d) correspond to the SM background processes contributing to the same final state.}
    \label{fig:diagrams}
\end{figure}

We study the effects of dimension-6 ($\kappa_{1}/\Lambda^2=\kappa_{2}/\Lambda^2$) and  dimension-8 ($\alpha_{8}/\Lambda^4$) 
anomalous couplings through the signal $e^{+}e^{-}\to Z \to \nu\bar{\nu}\gamma$ and relevant background processes in details by including parton showering and the response of detector effects at the Tera-Z phase for the FCC-ee and the CEPC.
In order to analyze the $e^+e^- \to \nu \bar{\nu}\gamma$ process, we consider the representative Feynman diagrams shown in Fig.~\ref{fig:diagrams}. Diagram~(a) corresponds to the signal process, where the $s$-channel $Z$-boson exchange includes potential contributions from new physics operators, with the photon emitted from the final-state neutrino line. In contrast, diagrams~(b)–(d) represent the SM contributions leading to the same final state. Specifically, diagram~(b) depicts $Z$-boson exchange with photon emission from the initial-state electron, while diagrams~(c) and (d) arise from $W$-boson exchange, with the photon radiated from the external $W$ boson or from the internal $W$ propagator, respectively. 
\begin{figure}[htbp]
\includegraphics[scale=0.5]{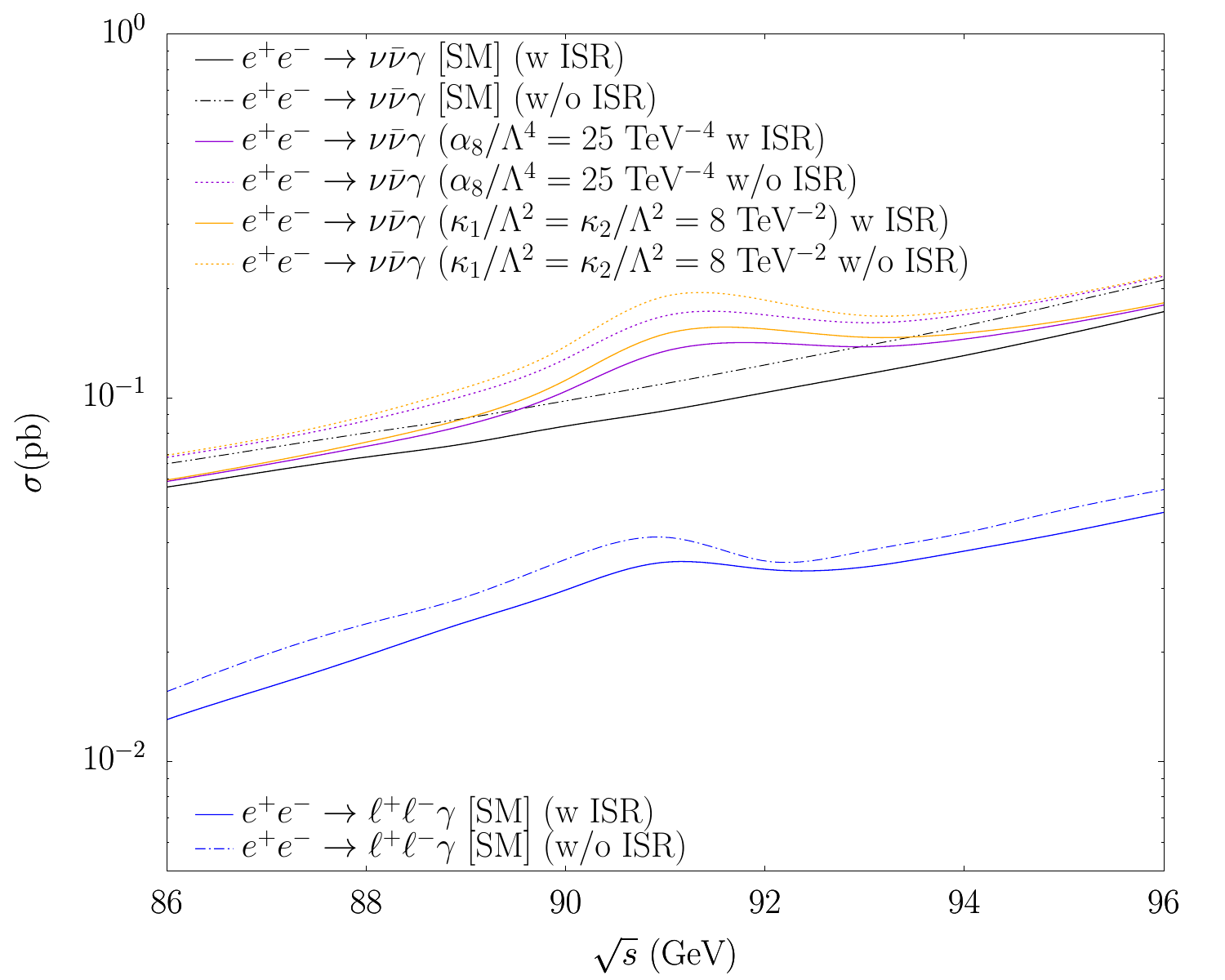}
\caption{The cross section for the process $e^{+}e^{-}\!\to\nu\bar{\nu}\gamma$ as a function of the 
center–of–mass energy $\sqrt{s}$, shown for the SM prediction and for two benchmark 
EFT scenarios.  The curves labeled “with ISR” and 
“without ISR’’ illustrate the impact of initial–state radiation on the line shape. \label{fig:xs_graphs}}
\end{figure}

The dominant SM background to the mono-photon signature arises from the irreducible process $e^+e^- \to \nu\bar{\nu}\gamma$, in which the neutrinos escape detection, leaving a single energetic photon in the final state. In addition, several reducible backgrounds can mimic the mono-photon signal when some of the final-state particles are not reconstructed by the detector. In particular, the process $e^+e^- \to \gamma\gamma\gamma$ contributes when two of the photons are either too soft, too collinear, or emitted in the forward/backward regions and therefore fall outside the detector acceptance, leaving only one observable photon in the fiducial region. Similarly, the process $e^+e^- \to \ell^+\ell^-\gamma$ can act as a background when the charged leptons are produced at very small polar angles and escape detection, while the radiated photon satisfies the mono-photon selection criteria.
These reducible backgrounds can be efficiently suppressed by imposing suitable fiducial and kinematic requirements on the observed photon, such as a minimum transverse momentum or energy and a restriction on its pseudorapidity.

Since initial-state radiation (ISR) can significantly modify production cross sections in the vicinity of a resonance, ISR effects are incorporated in our analysis using the \texttt{MadGraph5\_aMC@NLO v3.5.6}. In Fig.~\ref{fig:xs_graphs}, we present the generator level total cross sections of the signal and the relevant SM background processes as a function of the center-of-mass energy after applying a set of fiducial cuts that define mono-photon signature  by requiring a single photon in the central detector region with $p_T^{\gamma} > 10~\text{GeV}$ and $|\eta^{\gamma}| < 2.5$ while charged leptons are restricted to the forward and backward regions, $|\eta^{\ell^\pm}| > 2.5$ so that the process $e^+e^- \to \ell^+\ell^-\gamma$ contributes to the background only when the charged leptons escape detection. With these requirements, the contribution from the process $e^+e^- \to \ell^+\ell^-\gamma$ is strongly reduced. The minimum distance between photon (lepton) and lepton is also required to satisfy $\Delta R(\gamma(\ell),\ell)= \left[(\Delta\phi_{\gamma(\ell),\ell}])^2+(\Delta\eta_{\gamma(\ell),\ell}])^2\right]^{1/2} > 0.4$ where $\Delta\phi_{\gamma(\ell),\ell}$  and $\Delta\eta_{\gamma(\ell),\ell}$ are azimuthal angle and the pseudo rapidity difference between any two photons(leptons). Soft and collinear enhancements associated with photon radiation are regulated by the minimum transverse-momentum requirement and pseudo-rapidity on the observed photon, while additional QED radiation is treated inclusively via initial-state radiation. The solid curves correspond to the total cross sections of signal and relevant SM background processes without ISR, whereas the dashed curves indicate the results obtained when ISR effects are taken into account. The SM predictions are compared with EFT scenarios introducing anomalous couplings $\kappa_1/\Lambda^2 , \kappa_2/\Lambda^2$ and $\alpha_8/\Lambda^4$, with benchmark values $\kappa_1/\Lambda^2=\kappa_2/\Lambda^2 = 8~\text{TeV}^{-2}$ and $\alpha_8/\Lambda^4 = 25~\text{TeV}^{-4}$. 
It is important to clarify that the standard model does not contain a resonant $Z\to\nu\bar{\nu}\gamma$ process at tree level. The sharp peak observed in the $e^{+}e^{-}\!\to\nu\bar{\nu}\gamma$ cross section near $\sqrt{s}\simeq 91.2$~GeV in our figures is therefore not a standard model resonance associated with a physical $Z\to\nu\bar{\nu}\gamma$ decay. Instead, this peak arises from the effective $Z\nu\bar{\nu}\gamma$ vertex generated by the dimension–6 and dimension–8 operators. Once these operators are present, the process obtains an $s$–channel structure that mimics a resonance-like enhancement at the $Z$-boson mass. The SM contribution away from these EFT-induced effects remains smooth and non-resonant. Importantly, small deviations in this region can provide a sensitive probe of $\kappa_1/\Lambda^2=\kappa_2/\Lambda^2$ and $\alpha_8/\Lambda^4$ anomalous couplings. Consequently, the resonance region offers the highest discriminating power in separating potential signal effects from the irreducible background.

\begin{figure}[htbp]
    \centering
   \includegraphics[scale=0.4]{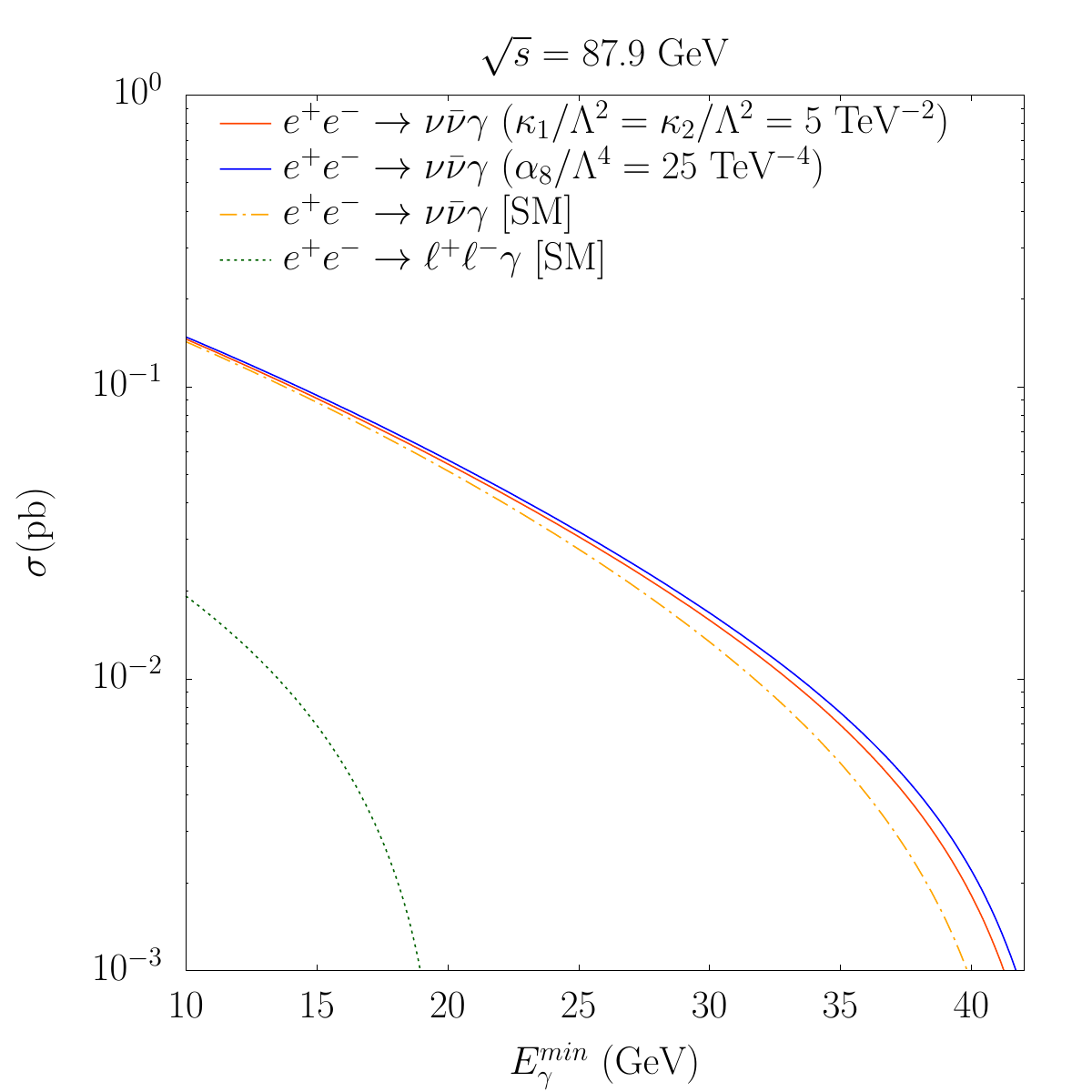}
\includegraphics[scale=0.4]{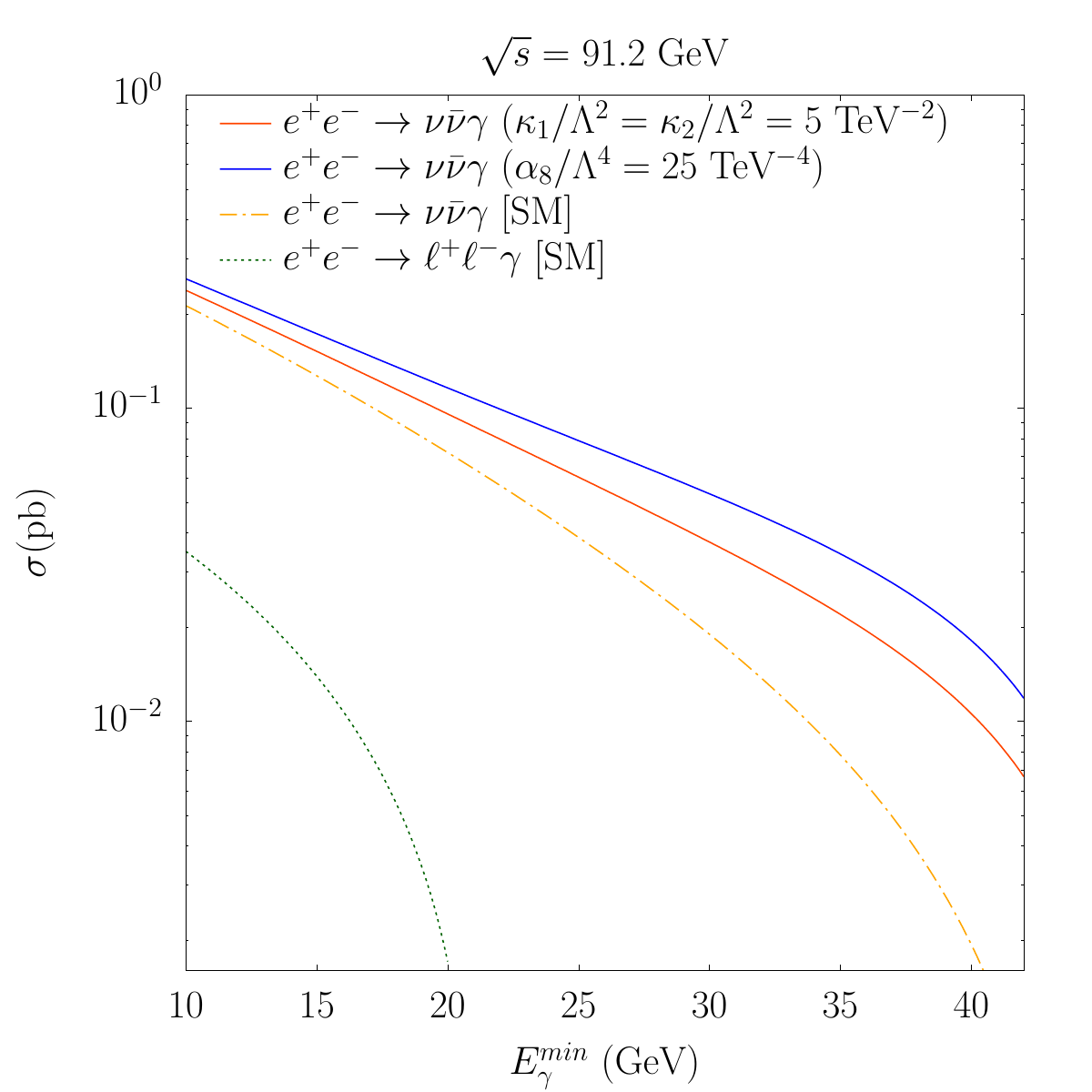}\\
\includegraphics[scale=0.4]{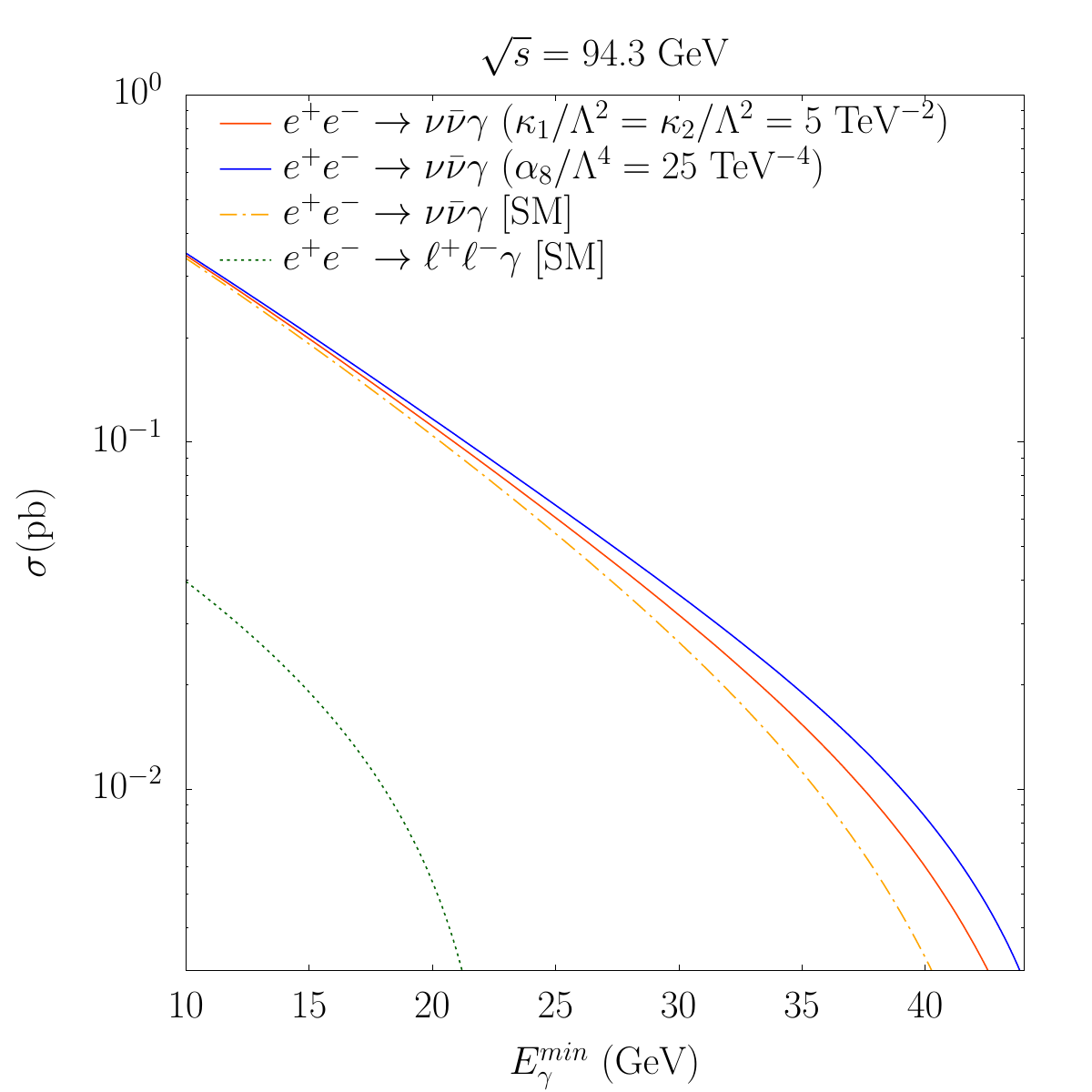}
    \caption{The total cross sections as a function of minimum photon energy cut $E_\gamma^{\rm min}$ for center-of-mass energies $\sqrt{s} = 87.7$, $91.2$, and $94.3$~GeV. The SM processes ($e^+e^- \to \nu\bar{\nu}\gamma$, and $e^+e^- \to \ell^+\ell^-\gamma$) are compared with contributions from effective field theory operators involving anomalous couplings.}
    \label{fig:crosssections}
\end{figure}

The minimum photon energy cut is also a sensitive probe of both SM dynamics and possible new physics, since different kinematic regions are shaped by infrared behavior, resonant enhancements, and gauge cancellations. In Fig.~\ref{fig:crosssections}, we show the total cross sections of the signal and relevant Standard Model background processes as functions of the minimum photon energy cut $E_\gamma^{\rm min}$ for three representative center-of-mass energies of the FCC-ee, namely $\sqrt{s} = 87.7$, $91.2$, and $94.3$~GeV.
The mono-photon signature is defined by requiring a single photon in the central detector region, while charged leptons are  outside the detector acceptance. With this fiducial selection, the irreducible background from $e^+e^- \to \nu\bar{\nu}\gamma$ dominates the low-$E_\gamma^{\rm min}$ region.
The contributions from  $\kappa_1/\Lambda^2=\kappa_2/\Lambda^2$ and $\alpha_8/\Lambda^4$ anomalous couplings, yield cross sections comparable to the background in the lower photon energy region. In particular, the benchmark values $\kappa_1/\Lambda^2=\kappa_2/\Lambda^2= 5~\text{TeV}^{-2}$ and $\alpha_8/\Lambda^4 = 25~\text{TeV}^{-4}$ have been chosen to illustrate the sensitivity reach in the signal channels. Notably, around the $Z$-boson resonance ($\sqrt{s} = 91.2$~GeV), the effects of new physics become more pronounced over the background, indicating a potential experimental sensitivity. This highlights the relevance of the LEP precision data and suggests that future $e^+e^-$ colliders could provide stringent tests of anomalous couplings. The contribution from the $e^+e^- \to \gamma\gamma\gamma$ process is not shown in Fig.~\ref{fig:xs_graphs} and Fig.~\ref{fig:crosssections}, since at the generator level it cannot consistently satisfy the mono-photon fiducial selection. Its potential impact is instead examined in the subsequent detector-level analysis, where realistic acceptance effects are taken into account.

We now proceed to a more detailed analysis of the mono-photon signature at the Tera-$Z$ stage of the FCC-ee, focusing on the process $e^{+}e^{-} \to \nu\bar{\nu}\gamma$ in the presence of anomalous $Z\nu\bar{\nu}\gamma$ interactions. The study is performed at three center-of-mass energies, $\sqrt{s}=87.9$, $91.2$, and $94.3~\text{GeV}$. In this stage of the analysis, event-level selections are applied in order to model realistic detector acceptance and to enforce a genuine mono-photon topology. In particular, we require exactly one photon to be reconstructed within the central detector region, accompanied by large missing transverse energy arising from invisible final-state particles. Additional photons or charged leptons are allowed only if they are emitted in the forward or backward regions and thus escape detection. This framework allows us to consistently assess the residual impact of reducible backgrounds, including multi-photon final states, once detector effects and realistic fiducial selections are taken into account.

For this purpose, samples with approximately one million events are generated for all relevant backgrounds and signals with seven benchmark values of $\kappa_1/\Lambda^2=\kappa_2/\Lambda^2$ ($\alpha_8/\Lambda^4$) coupling ranging between 1 TeV$^{-2}$ (5 TeV$^{-4}$) and 10 TeV$^{-2}$ (40 TeV$^{-4}$) at three center-of-mass energies of the Tera-Z stage of the FCC-ee. All generated signal and relevant background events includes ISR effects and basic generated level kinematic cuts; $p_{T}^\gamma > 10~\text{GeV}$, $|\eta^{\gamma}| < 2.5$,   $|\eta^{\ell}| > 2.5$ and $\Delta R(\gamma(\ell),\ell)>0.4$. Parton showering and hadronization are carried out using the Pythia 8.2 package \cite{Sjostrand:2014zea}. Subsequently, the events are interfaced with Delphes 3.4.2 \cite{deFavereau:2013fsa} software to model the response of the corresponding detector in the form of resolution functions and efficiencies for the FCC-ee. Latest Innovative Detector for Electron–Positron Accelerators (IDEA) card (delphes\_card\_IDEA.tcl) is chosen for the FCC-ee/CEPC detector concept.

The analysis is initiated by requiring at least one reconstructed photon in the final state ($N_\gamma > 0$). 
An angular requirement of $ 10^\circ < \theta_\gamma < 170^\circ$ is also imposed on the highest $p_T$ photon, considering the geometric acceptance of realistic $e^{+}e^{-}$ collider detectors. Additionally, generator level cuts are applied throughout the subsequent analysis to ensure a uniform treatment of detector acceptance effects. 

The analysis is initiated by requiring at least one reconstructed photon in the final state ($N_\gamma > 0$). An angular requirement of $10^\circ < \theta_\gamma < 170^\circ$  as well as $|\eta^{\gamma_1}| < 2.5$ is imposed on the highest-$p_T$ photon ($\gamma_1$) with $p_T^{\gamma_1}>10$ GeV, reflecting the geometric acceptance of a realistic $e^{+}e^{-}$ collider detector (define the central detector region). Additional photons and leptons, when present, are allowed only if they are emitted in the forward or backward regions, $|\eta^{\gamma_2,\gamma_3}| > 2.5$ and $|\eta^{\ell_1,\ell_2}| > 2.5$, and therefore escape detection. These selections ensure that multi-photon background $e^{+}e^{-} \to \gamma\gamma\gamma$ and  $e^{+}e^{-} \to \ell^+\ell^-\gamma$ background contribute only when they mimic a mono-photon signature due to detector acceptance effects.
\begin{figure}[htbp]
\includegraphics[scale=0.40]{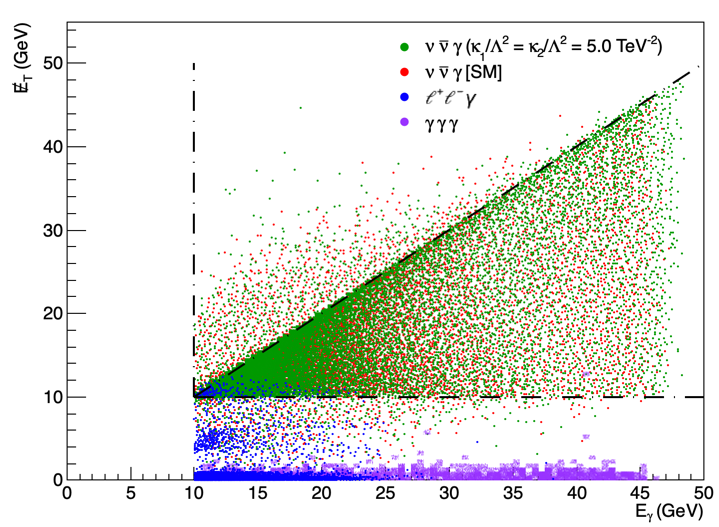}
 \caption{
 Normalized distribution of the $(E_{\gamma_1}, \slashed{E}_T)$ plane for the signal process $e^+e^- \to \nu\bar{\nu}\gamma$ with anomalous couplings $\kappa_1/\Lambda^2=\kappa_2/\Lambda^2=5.0~\text{TeV}^{-4}$ (Green points), the SM backgrounds $e^+e^- \to \nu\bar{\nu}\gamma$ (Red points), $e^+e^- \to \ell^+\ell^-\gamma$ (Blue points) and  $e^+e^- \to \gamma\gamma\gamma$ (Purple points) for the FCC-ee at $\sqrt{s} = 91.2$ GeV. The black dashed lines indicate the kinematic cuts.
  \label{metvsegamma}}
\end{figure}
After applying the inclusive event selection, the two-dimensional distributions in the $E_\gamma$--$\slashed{E}_T$ plane shown in Fig.~\ref{metvsegamma} are obtained and used to motivate the subsequent kinematic requirements. The figure illustrates the case of anomalous couplings $\kappa_1/\Lambda^2=\kappa_2/\Lambda^2 = 5.0~\text{TeV}^{-2}$ and includes, in addition to the signal, all relevant Standard Model background contributions. The signal process $e^+e^- \to \nu\bar{\nu}\gamma$ with anomalous couplings (green) populates a broad region of the phase space, characterized by sizable missing transverse energy that is strongly correlated with the photon energy. The corresponding Standard Model contribution to $e^+e^- \to \nu\bar{\nu}\gamma$ (red) exhibits a similar kinematic correlation but with a reduced overall density, reflecting the absence of EFT-induced enhancements. The background process $e^+e^- \to \ell^+\ell^-\gamma$ (blue) is mainly concentrated at low values of $\slashed{E}_T$, as no genuine invisible particles are produced in this channel and the apparent missing energy arises predominantly from detector effects or limited acceptance. The $e^+e^- \to \gamma\gamma\gamma$ background (purple) is strongly suppressed in the high-$\slashed{E}_T$ region and accumulates near very small missing transverse energy, consistent with the fully visible nature of the final state when at least one photon is reconstructed in the central detector region. The dashed lines in Fig.~\ref{metvsegamma} indicate the applied kinematic requirements: photon transverse energy $E_{\gamma_1} > 10~\mathrm{GeV}$, missing transverse energy $\slashed{E}_T > 10~\mathrm{GeV}$, and the condition $E_T \leq E_{\gamma_1}$. These selections efficiently suppress reducible backgrounds while preserving the signal-rich region of the phase space. A comparison of similar distributions obtained for different values of $\kappa_1/\Lambda^2=\kappa_2/\Lambda^2$ and $\alpha_8/\Lambda^4$ shows that anomalous couplings primarily modify the population density and shape of the signal region, whereas the separation between signal and backgrounds remains robust.

\begin{figure}[htbp]
\includegraphics[scale=0.45]{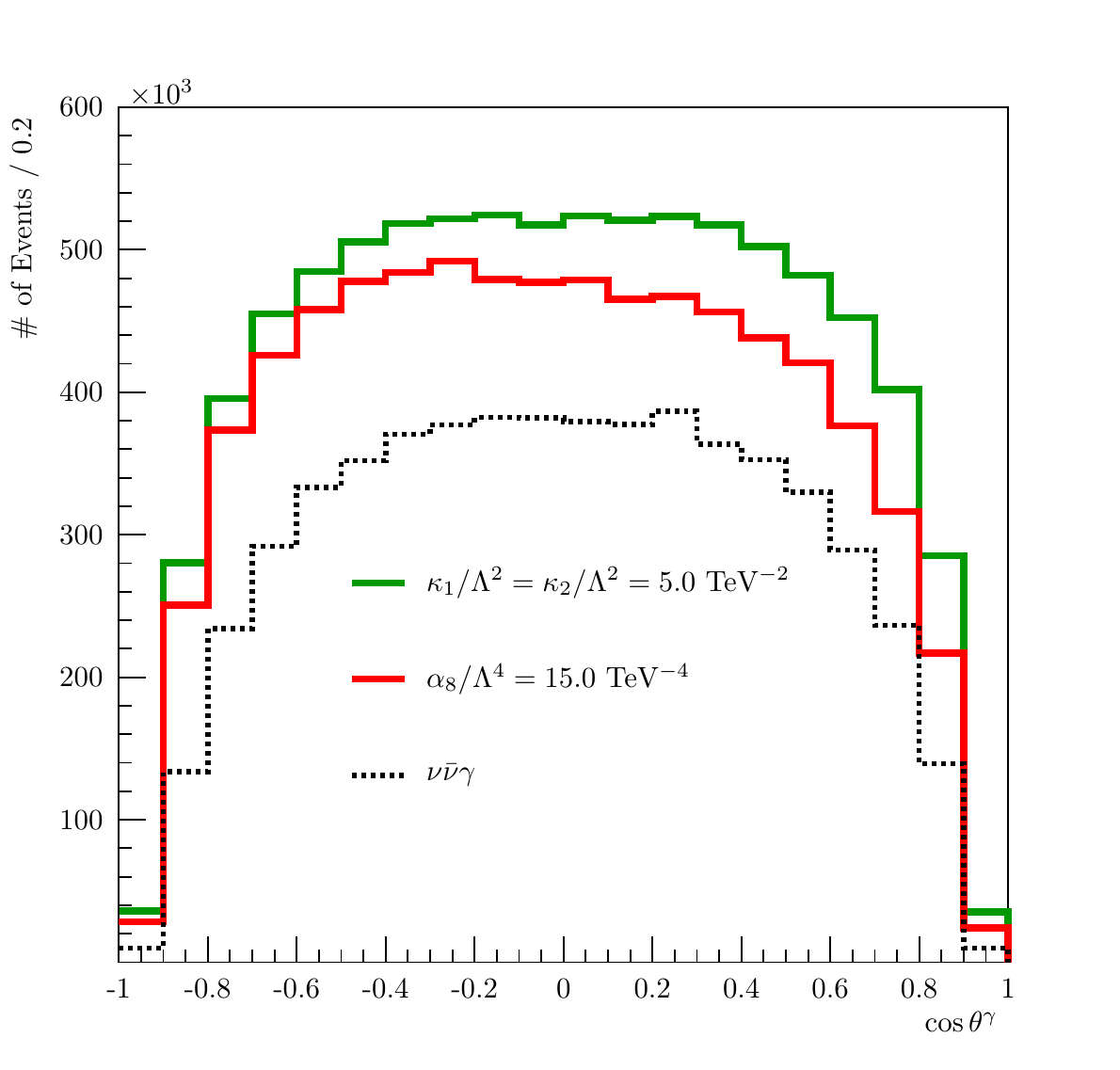}
 \caption{ Normalized $\cos\theta$ distribution of the leading final-state photon for the SM backgrounds and for representative  dimension-6 and dimension-8 effective operator scenarios, corresponding to $\kappa_1/\Lambda^2=\kappa_2/\Lambda^2 = 5.0~\mathrm{TeV}^{-2}$ and $\alpha_8/\Lambda^4 = 15.0~\mathrm{TeV}^{-4}$ for the FCC-ee at $\sqrt{s} = 91.2$ GeV, respectively.\label{costheta}}
\end{figure}
\begin{figure}[htbp]
\includegraphics[scale=0.4]{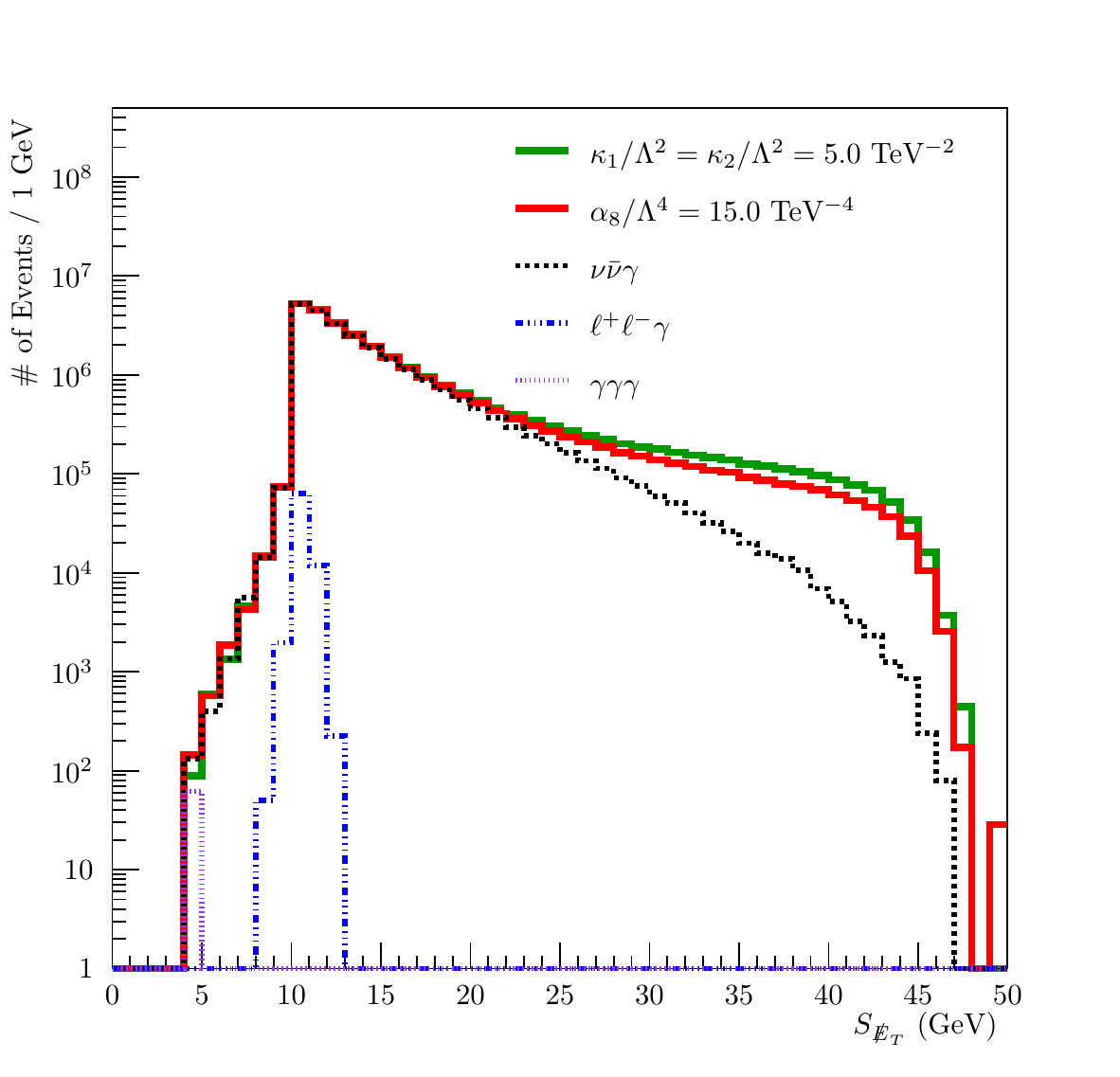}
\includegraphics[scale=0.4]{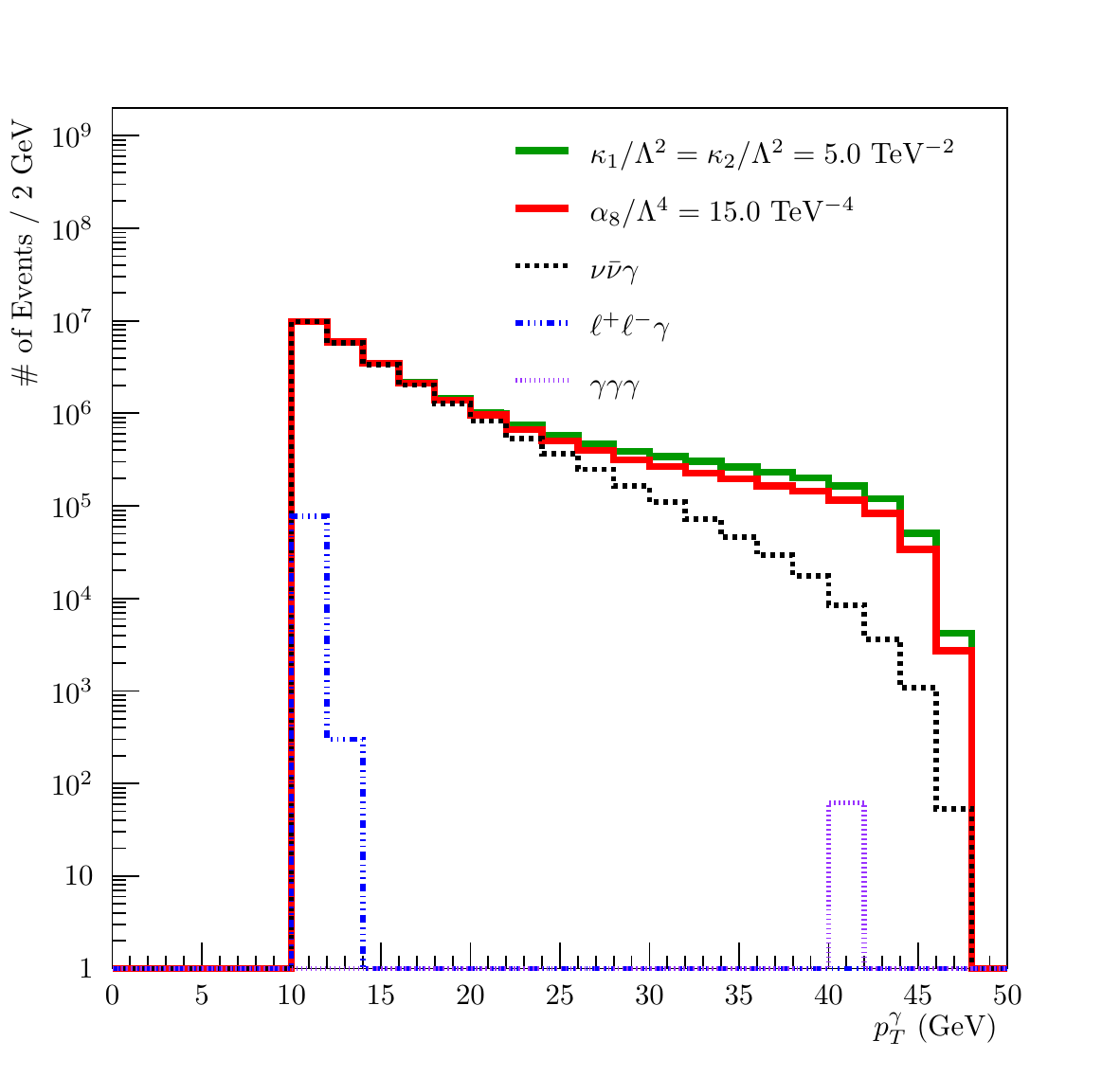}\\

\caption{The normalized distributions of missing transverse energy significance  ($S_{\slashed{E}_T}$) and the transverse momentum of photon ($p_{T}^{\gamma}$) for the signal with non-zero anomalous couplings ($\kappa_1/\Lambda^2=\kappa_2/\Lambda^2= 5~\mathrm{TeV}^{-2}$ and $\alpha_8/\Lambda^{4} = 15~\mathrm{TeV}^{-4}$) and relevant background processes after applying the selection cuts $E_{\gamma_1} > 10~\mathrm{GeV}$ and $\slashed{E}_T > 10~\mathrm{GeV}$ at the FCC-ee with $\sqrt{s} = 91.2$ GeV.
    \label{fig:sig_pt}}
\end{figure}

To assess the extent to which the  dimension-6 and dimension-8 effective operator scenarios can be experimentally disentangled, we investigate kinematic shape information encoded in angular observables of the final-state photon, shown in Fig.~\ref{costheta}. The figure presents the $\cos\theta_\gamma$, the angle $\theta_\gamma$ between the photon and beam line, distribution for the standard model backgrounds and signals with the dimension-6 operator characterized by $\kappa_1/\Lambda^2=\kappa_2/\Lambda^2 = 5.0~\mathrm{TeV}^{-2}$, and the dimension-eight operator with $\alpha_8/\Lambda^4 = 15.0~\mathrm{TeV}^{-4}$. Owing to their distinct Lorentz structures and energy-scaling behavior within the effective field theory framework, these operators generate qualitatively different angular patterns. In particular, the contribution induced by the dimension-8 operator leads to a more pronounced forward--backward asymmetry in the $\cos\theta^{\gamma}$ distribution when compared to both the standard model and the dimension-6 scenario. This behavior originates from the different Lorentz and tensor structures of the higher-dimensional operators, which modify the helicity composition of the amplitude and its angular dependence. In contrast to the dimension-6 interaction, the dimension-8 operator introduces additional momentum contractions that enhance asymmetric angular configurations. 
As a consequence of its Lorentz structure, the dimension-eight operator induces a nontrivial forward--backward asymmetry in the photon angular distribution, characterized by an uneven population of positive and negative $\cos\theta^\gamma$ regions. This asymmetry can be exploited as a discriminating observable between the two effective operator hypotheses.
Fig.~\ref{fig:sig_pt} presents the distributions of two complementary kinematic variables, the photon transverse momentum ($p_T^{\gamma_1}$) and the missing transverse energy significance ($S_{\slashed{E}_T} = \slashed{E}_T^2 / E_T^{\gamma_1}$), which are used to discriminate the signal process $e^+e^- \to \nu\bar{\nu}\gamma$ from the relevant SM backgrounds. The left panel shows the $S_{\slashed{E}_T}$ distribution, which is particularly sensitive to the presence of genuine invisible particles in the final state. The signal and the SM $\nu\bar{\nu}\gamma$ background populate the high-$S_{\slashed{E}_T}$ region, reflecting the presence of real missing transverse energy carried away by neutrinos. 
In contrast, the reducible backgrounds $e^+e^- \to \ell^+\ell^-\gamma$ and $e^+e^- \to \gamma\gamma\gamma$ are strongly concentrated at low $S_{\slashed{E}_T}$ values, since these processes do not involve true invisible particles and any apparent missing energy arises primarily from detector resolution effects or particles escaping the fiducial acceptance. Deviation between the signal and the SM $\nu\bar{\nu}\gamma$ background starts to be seen at approximately 16. Therefore, events with $S_{\slashed{E}_T} > 16$ are selected for the subsequent analysis. The right panel of Fig.~\ref{fig:sig_pt} shows the transverse-momentum distribution of the photon.
The irreducible SM background $e^+e^- \to \nu\bar{\nu}\gamma$ decreases rapidly with increasing $p_T^{\gamma_1}$ and dominates the low-$p_T^{\gamma_1}$ region. The $e^+e^- \to \ell^+\ell^-\gamma$ background exhibits a harder spectrum at intermediate
$p_T^{\gamma}$, while the residual contribution from the $e^+e^- \to \gamma\gamma\gamma$ process remains suppressed over the entire range after the mono-photon selection.
The EFT-induced signal distributions follow the SM prediction at low transverse momentum but show a relative enhancement at intermediate-to-high $p_T^{\gamma}$ values, where the SM backgrounds are already significantly reduced.
The details of the cuts and their efficiencies are further quantified in the cut-flow table (Table~\ref{tab:cutflow}) which demonstrates the number of simulated events (normalized to  $\mathcal{L}_{\text{int}}=125~\text{ab}^{-1}$ for $\sqrt{s} = 91.2$ GeV) after each selection step for the  $e^+e^- \to \nu\bar{\nu}\gamma$ process with anomalous couplings $\kappa_1/\Lambda^2=\kappa_2/\Lambda^2= 5.0~\text{TeV}^{-2}$ and $\alpha_8/\Lambda^4 = 15~\text{TeV}^{-4}$, as well as for the SM $\nu\bar\nu\gamma$, $\ell^+\ell^-\gamma$ and $\gamma\gamma\gamma$ backgrounds.
\begin{table}
\caption{The normalized number of events ($\sqrt{s} = 91.2$ GeV and $\mathcal{L}_{\text{int}}=125~\text{ab}^{-1}$) surviving after each selection for the 
$\nu\bar\nu\gamma$ process with anomalous couplings 
$\kappa_1/\Lambda^2=\kappa_2/\Lambda^2= 5.0~\text{TeV}^{-2}$ and 
$\alpha_8/\Lambda^{4} = 15~\mathrm{TeV}^{-4}$, together with the SM backgrounds; 
$\nu\bar\nu\gamma$, $\ell^+\ell^-\gamma$ and $\gamma\gamma\gamma$. Each row corresponds to the application of an additional requirement, 
illustrating the progressive reduction of events and the relative impact of the selection criteria on different processes.\label{tab:cutflow}}
\scriptsize{\begin{ruledtabular}
\begin{tabular}{l c c c c c}
Cuts&$\nu\bar\nu\gamma$ &$\nu\bar\nu\gamma$  &$\nu\bar\nu\gamma$& $\ell^+\ell^-\gamma$&$\gamma\gamma\gamma$ \\ 
& ($\kappa_1/\Lambda^2=\kappa_2/\Lambda^2= 5.0~\text{TeV}^{-2}$)& $(\alpha_8/\Lambda^4=15$ TeV$^{-4}$) &(SM)& (SM)& (SM)\\ \hline
$N_{\gamma}>0$ &2.92$\times 10^{7}$&2.82$\times 10^{7}$&2.61$\times 10^{7}$&6.16$\times 10^{7}$& 4.31$\times 10^{6}$ \\ \hline
$p_T^{\gamma_1}>10$ GeV, $|\eta^{\gamma_1}| < 2.5$, &&&&  \\ 
$\Delta R(\gamma(\ell),\ell)>0.4$,   & 2.82$\times 10^{7}$ & 2.72$\times 10^{7}$ & 2.51$\times 10^{7}$& 4.08$\times 10^{4}$&3.53$\times 10^{6}$  \\ 
$ 10^\circ<\theta_\gamma<170^\circ$ & &&&  \\ 
$|\eta^{\ell_1,\ell_2}| > 2.5$, $|\eta^{\gamma_2,\gamma_3}| > 2.5$&&&  \\ \hline
 $E_{\gamma}>$10 GeV,  &&&&&\\
$\slashed{E}_T >$ 10 GeV,  &2.78$\times 10^{7}$&2.68$\times 10^{7}$ & 2.47$\times 10^{7}$ &6.18$\times 10^{1}$&7.72$\times 10^{4}$ \\ 
 $\slashed{E}_T\leq E_{\gamma}$  &&&&& \\ \hline
$S_{\slashed{E}_T}>16$& 8.48$\times 10^{6}$ & 7.61$\times 10^{6}$ & 5.73$\times 10^{6}$& 0 & 0 \\ 
\end{tabular}
\end{ruledtabular}}
\end{table}
Each row corresponds to the cumulative application of the listed selection criteria. The basic photon kinematic requirements, $p_T^{\gamma_1}>10$~GeV and $|\eta^{\gamma_1}|<2.5$, together with the isolation condition $\Delta R(\gamma(\ell),\ell)>0.4$ and the angular acceptance cut $10^\circ<\theta_\gamma<170^\circ$, already reduce the
$e^+e^-\to\ell^+\ell^-\gamma$ background significantly.
The additional requirement that charged leptons be emitted in the forward or backward regions, $|\eta^{\ell_1,\ell_2}|>2.5$, further suppresses this background by vetoing events in which leptons are detectable in the central region. The subsequent missing-energy-related selections,
$E_\gamma>10$~GeV, $\slashed{E}_T>10$~GeV, and $\slashed{E}_T\leq E_\gamma$,
efficiently reduce backgrounds without genuine invisible particles. In particular, the $e^+e^-\to\gamma\gamma\gamma$ background is strongly suppressed at this stage, since any apparent missing transverse energy arises only from detector effects or photons escaping
the acceptance. Finally, the requirement $S_{\slashed{E}_T}>16$ provides a powerful discrimination between signal-like events with genuine missing energy and reducible backgrounds. After this cut, both the $e^+e^-\to\ell^+\ell^-\gamma$ and $e^+e^-\to\gamma\gamma\gamma$
backgrounds are completely eliminated, while a sizable fraction of the signal events remains.The SM $\nu\bar{\nu}\gamma$ background is reduced by approximately one order of magnitude, whereas the signal retains a significantly higher efficiency. This demonstrates that the applied selection strategy is highly effective in isolating
mono-photon events arising from anomalous $Z\nu\bar{\nu}\gamma$ interactions.
\section{Results on the anomalous $Z\nu\bar{\nu}\gamma$ couplings}
In this section, we present the limits on the anomalous $Z\nu\bar{\nu}\gamma$ couplings, specifically $\kappa_1/\Lambda^2 = \kappa_2/\Lambda^2$ and $\alpha_8/\Lambda^{4}$, obtained from the analysis of the $e^{+}e^{-} \to \nu\bar{\nu}\gamma$ process at the FCC-ee and CEPC. The expression used to determine the median expected statistical significance ($SS$) for discovery of new phenomena, incorporating systematic uncertainties in high-energy physics, is given as follows \cite{Cowan:2010js} :
\begin{eqnarray}
SS = \sqrt{2\left[(S+B)\ln\left(\frac{(S+B)(1+\delta^2B)}{B+\delta^2B(S+B)}\right) - \frac{1}{\delta^2}\ln\left(1+\delta^2\frac{S}{1+\delta^2B}\right)\right]} 
\label{SS}\end{eqnarray}
where $S$ and $B$ are the total normalized number of signal and SM background events obtained by integrating the transverse momentum distribution of the photon after applying all mentioned cuts in Table \ref{tab:cutflow} and $\delta$ indicates the systematic uncertainty. In the limit $\delta \to 0$, Eq.~(\ref{SS}) becomes :
\begin{eqnarray}
SS = \sqrt{2\left[(S+B)\ln\left(1+S/B\right) - S\right]} \quad.
\end{eqnarray}

The FCC-ee is assumed to operate at center-of-mass energies of $\sqrt{s}=87.9$~GeV with an integrated luminosity of $\mathcal{L}_{\text{int}} = 40~\text{ab}^{-1}$, $\sqrt{s}=91.2$~GeV with $\mathcal{L}_{\text{int}} = 125~\text{ab}^{-1}$, and $\sqrt{s}=94.3$~GeV with $\mathcal{L}_{\text{int}} = 40~\text{ab}^{-1}$. The statistical significances obtained at each energy point are subsequently combined to yield the total signal significance, defined as
\begin{equation}
SS_{\text{tot}} = \sqrt{ \left( SS_{87.9~\text{GeV}} \right)^2 + \left( SS_{91.2~\text{GeV}} \right)^2 + \left( SS_{94.3~\text{GeV}} \right)^2 } .
\end{equation}

\begin{figure}[httb!]
\includegraphics[scale=0.4]{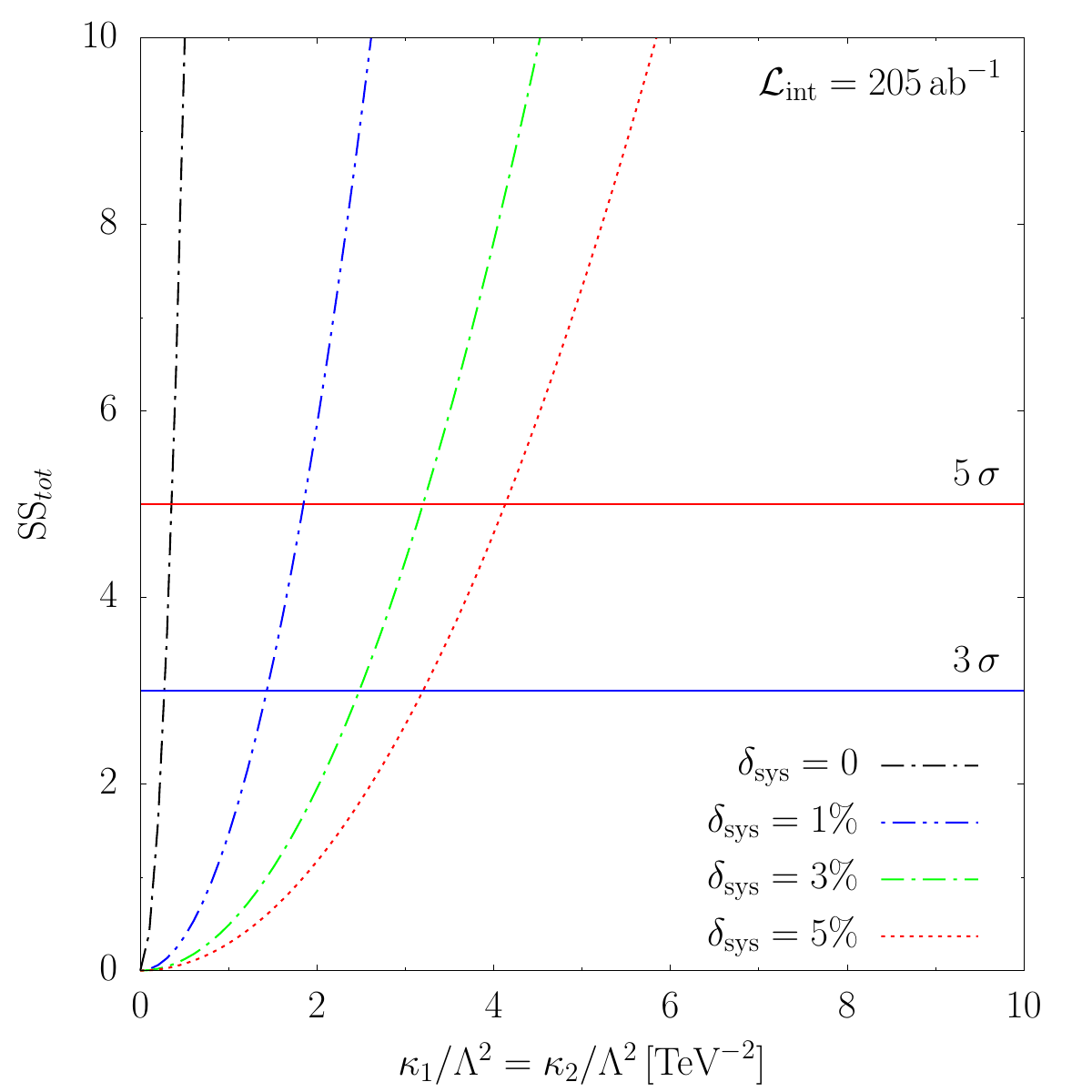}
\includegraphics[scale=0.4]{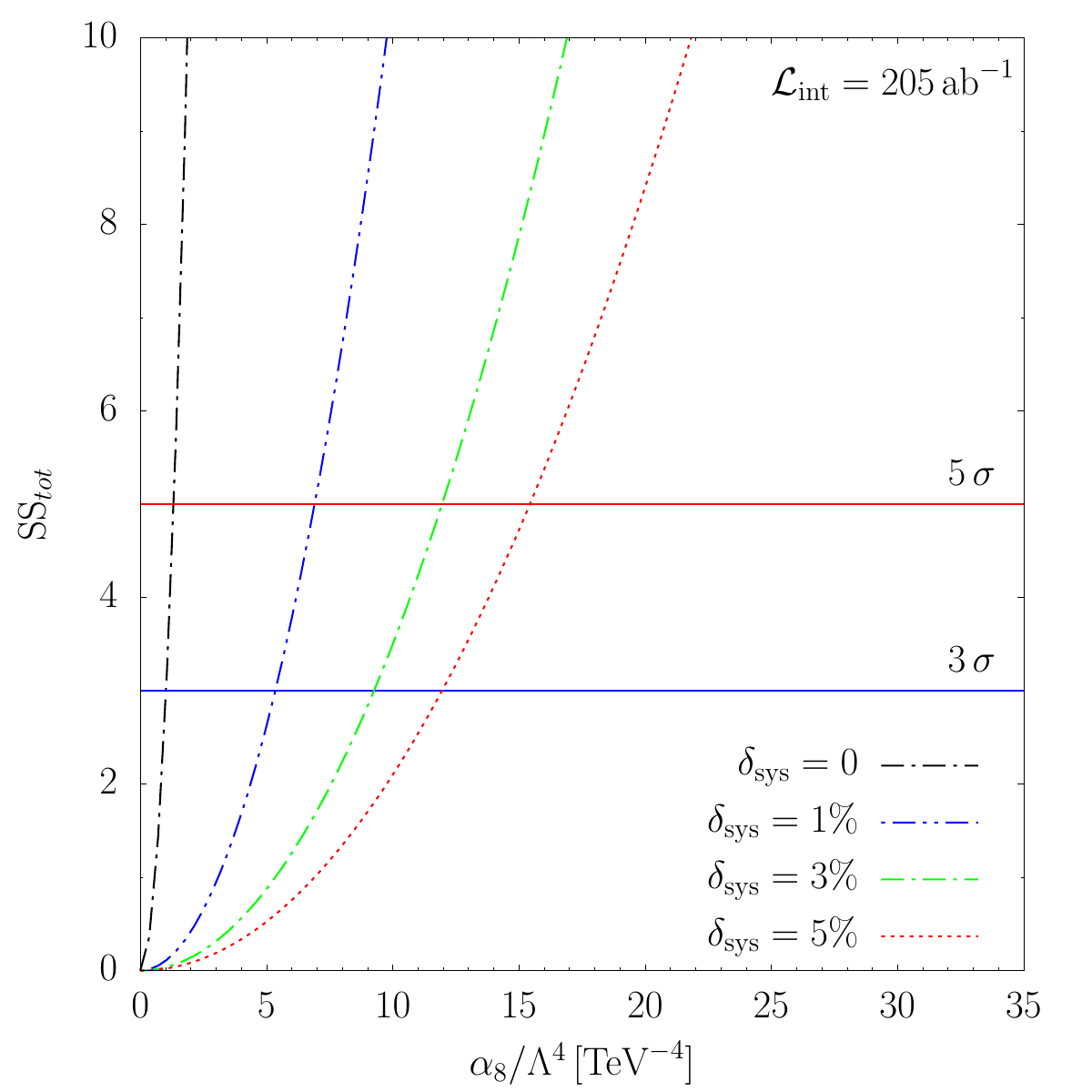}
\caption{ The total Statistical Significance as a function of the anomalous couplings $\kappa_1/\Lambda^2$ = $\kappa_2/\Lambda^2$ (left) and $\alpha_8/\Lambda^{4}$ (right) at $\sqrt{s}=91.2$ GeV with $\mathcal{L}_{\text{int}}=205~\text{ab}^{-1}$. Curves correspond to different systematic uncertainties ($\delta_{\text{sys}}=0$--$5\%$). Horizontal lines indicate the $3\sigma$ and $5\sigma$ thresholds.
 \label{fig:ss}}
\end{figure}

Fig.~\ref{fig:ss} shows the total statistical significance ($SS_{tot}$) as a function of the anomalous couplings $\kappa_1/\Lambda^2$=$\kappa_2/\Lambda^2$ (left) and $\alpha_8/\Lambda^{4}$ (right) for the combined integrated luminosity of $\mathcal{L}_{\text{int}} = 205~\text{ab}^{-1}$. The curves correspond to different assumptions on the total systematic uncertainties, $\delta_{\text{sys}}$, ranging from 0\% to 5\%, while the horizontal lines at $3\sigma$ and $5\sigma$ indicate the conventional thresholds for evidence and discovery, respectively. The expected limits on the anomalous couplings for $\kappa_1/\Lambda^2=\kappa_2/\Lambda^2$ and $\alpha_8/\Lambda^{4}$ are derived from the intersection points of the significance curves with the $3\sigma$ and $5\sigma$ reference lines. In the ideal case of $\delta_{\text{sys}} = 0$, the collider exhibits excellent discovery potential, reaching the $5\sigma$ significance level for $\kappa_1/\Lambda^2$=$\kappa_2/\Lambda^2$ values below $1~\text{TeV}^{-2}$ and $\alpha_8/\Lambda^4$ values of order $10~\text{TeV}^{-4}$. However, as the systematic uncertainty increases, the statistical significance decreases noticeably, with $\delta_{\text{sys}} = 5\%$ reducing the sensitivity by roughly a factor of two. 
Building upon these sensitivity estimates, the corresponding upper limits on the branching ratio $\mathrm{BR}(Z \to \nu\bar{\nu}\gamma)$ are summarized in Table~\ref{limits_br}. The quoted FCC-ee limits are obtained by statistically combining the results from the three center-of-mass energies, $\sqrt{s}=87.9$, $91.2$, and $94.3~\mathrm{GeV}$, corresponding to a total integrated luminosity of $\mathcal{L}_{\text{int}}=205~\mathrm{ab}^{-1}$. For comparison, the values shown in parentheses represent the projected sensitivities for the CEPC operating at $\sqrt{s}=91.2~\mathrm{GeV}$ with an integrated luminosity of $100~\mathrm{ab}^{-1}$.

In the absence of systematic uncertainties ($\delta_{\text{sys}}=0\%$), the branching ratio associated with the coupling combination $\kappa_1/\Lambda^2 = \kappa_2/\Lambda^2$ can be constrained down to $1.90\times10^{-9}$ ($2.16\times10^{-9}$) at the $3\sigma$ level and $3.18\times10^{-9}$ ($3.61\times10^{-9}$) at the $5\sigma$ level for the FCC-ee (CEPC). As the systematic uncertainty increases to $5\%$, the corresponding limits deteriorate substantially to $2.56\times10^{-7}$ ($2.68\times10^{-7}$) and $4.27\times10^{-7}$ ($4.46\times10^{-7}$) at the $3\sigma$ and $5\sigma$ confidence levels, respectively.

A similar dependence on systematic effects is observed for the $\alpha_8/\Lambda^4$ operator. For $\delta_{\text{sys}}=0\%$, the combined FCC-ee (CEPC) limits reach $1.95\times10^{-9}$ ($2.22\times10^{-9}$) at the $3\sigma$ level and $3.27\times10^{-9}$ ($3.71\times10^{-9}$) at the $5\sigma$ level, while they weaken to $2.67\times10^{-7}$ ($2.80\times10^{-7}$) and $4.46\times10^{-7}$ ($4.67\times10^{-7}$) when $\delta_{\text{sys}}$ is increased to $5\%$. This pronounced sensitivity to systematic uncertainties highlights the crucial importance of stringent experimental control, particularly in the determination of the integrated luminosity, photon energy calibration, and background normalization. Nevertheless, even under conservative assumptions, the combined FCC-ee sensitivity at the $Z$ pole surpasses existing experimental limits by several orders of magnitude, emphasizing its strong potential to probe anomalous interactions through the radiative $Z \to \nu\bar{\nu}\gamma$ decay channel.

\begin{table}[httb!]
\centering
\caption{The limits on BR($Z\to\nu\bar\nu\gamma$) at $3\sigma$ and $5\sigma$, considering $\delta_{\text{sys}} = 0\%$, $1\%$, $3\%$ and $5\%$ systematic uncertainties for $\mathcal{L}_{int} = 205 $ ab$^{-1}$ at the FCC-ee with combined $\sqrt s$= 87.9 GeV, 91.2 GeV and 94.3 GeV. The limits for the CEPC at $\sqrt s$=91.2 GeV with an integrated luminosity of 100 ab$^{-1}$ are given in the parenthesis.\label{limits_br}}
\begin{ruledtabular}
\begin{tabular}{c c c c }
& & \multicolumn{2}{c}{ BR($Z\to\nu\bar\nu\gamma$) } \\ \cline{3-4} 
Coefficient  & $\delta_{\text
{sys}}$ & $3\sigma$ & $5\sigma$  \\ \hline
\multirow{4}{*}{$\kappa_1/\Lambda^2$=$\kappa_2/\Lambda^2$} & 0\% & 1.90$\times 10^{-9}$ (2.16$\times 10^{-9}$)& 3.18$\times 10^{-9}$(3.61$\times 10^{-9}$)  \\ 
 & 1\% & 5.12$\times 10^{-8}$ (5.38$\times 10^{-8}$)  & 8.54$\times 10^{-8}$(8.94$\times 10^{-8}$)  \\ 
  & 3\% &1.54$\times 10^{-7}$ (1.61$\times 10^{-7}$) & 2.56$\times 10^{-7}$ (2.68$\times 10^{-7}$)  \\ 
 & 5\% & 2.56$\times 10^{-7}$ (2.68$\times 10^{-7}$) & 4.27$\times 10^{-7}$ (4.46$\times 10^{-7}$) \\ \hline
\multirow{4}{*}{$\alpha_8/\Lambda^{4}$} & 0\% & 1.95$\times 10^{-9}$ (2.22$\times 10^{-9}$) & 3.27$\times 10^{-9}$ (3.71$\times 10^{-9}$)  \\
 & 1\% & 5.35$\times 10^{-8}$ (5.63$\times 10^{-8}$) & 8.91$\times 10^{-8}$ (9.36$\times 10^{-8}$)\\
 & 3\% & 1.60$\times 10^{-7}$ (1.69$\times 10^{-7}$) & 2.68$\times 10^{-7}$ (2.80$\times 10^{-7}$)\\
 & 5\% & 2.67$\times 10^{-7}$ (2.80$\times 10^{-7}$) & 4.46$\times 10^{-7}$ (4.67$\times 10^{-7}$)
\end{tabular}
\end{ruledtabular}
\end{table}
\section{Conslusion}
Having an advantage of clean experimental environment and good knowledge of initial state, the FCC-ee and the CEPC  are two proposed future circular high energy electron-positron collider project with the objective of precise measurement of many reactions as well as exploring BSM. One of the proposed stages at different center-of-mass energies of both colliders is to  operate as high-luminosity Z-boson factories, which will allow for precise measurements of electroweak parameters and the exploration of BSM. Therefore, this gives an opportunity to investigate rare Z boson decays to probe hidden sectors, test theoretical models, and place constraints on new physics scenarios. It is anticipated that future experiments at high-luminosity colliders such as the FCC-ee and the CEPC will further explore these decays, potentially leading to significant breakthroughs in our understanding of fundamental physics. In this study, we have examined the contribution of dimension-6 and dimension-8 effective operators that determine the $Z\nu\bar{\nu}\gamma$ coupling to $e^{+}e^{-}\to Z \to \nu\bar{\nu}\gamma$ process. To optimize the signal selection and achieve stronger limits, we employed a cut-based analysis relying on key kinematic variables the photon energy ($E_\gamma$), missing transverse energy ($\slashed{E}_T$), and the missing transverse energy significance ($S_{\slashed{E}_T}$) which proved highly effective in isolating the signal from the dominant SM backgrounds. Our results without systematic uncertainty demonstrate that the estimated upper limits on the branching ratio of $Z\to\nu\nu\gamma$ decay reach the order of $10^{-9}$, representing an improvement of several orders of magnitude over the existing LEP constraints. Furthermore, even with a conservative systematic uncertainty of $5\%$, the proposed analysis remains highly competitive, still outperforming current experimental bounds by a wide margin. This underscores the importance of precise calibration, luminosity measurement, and background modeling in the experimental program. 

Due to the Lorentz structure of the underlying effective operators, the angular observables such as the angle between highest $p_T$ of the photon and beam line can play a crucial role to assess the extent to which the  dimension-6 and dimension-8 effective operator scenarios can be experimentally disentangled.

In light of these findings, we conclude that the Tera-Z program at the FCC-ee and CEPC will not only provide a stringent test of the SM at the loop level but also possess a remarkable discovery potential for BSM. 

\begin{acknowledgments}
The numerical calculations reported in this paper were partially performed at TUBITAK ULAKBIM, High Performance and Grid Computing Center (TRUBA resources).
\end{acknowledgments}

\end{document}